\documentclass[11pt]{article}
\usepackage[a4paper,margin=2.4cm]{geometry}
\usepackage{amsmath,amssymb,amsthm}
\usepackage{booktabs}
\usepackage{array}
\usepackage{graphicx}
\usepackage{siunitx}
\usepackage{xcolor}
\usepackage{tikz}
\usetikzlibrary{arrows.meta,positioning,fit}
\usepackage[section]{placeins}
\usepackage[colorlinks=true,linkcolor=blue!50!black,citecolor=blue!50!black,urlcolor=blue!50!black]{hyperref}
\title{Verifying the Outputs of PDE Solvers:\\
From Finite Elements to Neural Surrogates}
\author{Konstantinos Dovelos\\ \small Dovelos Engineering \\ \small \texttt{konstantinos.dovelos@gmail.com}}
\date{}

\begin{document}
\maketitle

\begin{abstract}
Physics simulation outputs increasingly come from solvers that no expert inspects,
such as third-party packages in automated pipelines, programs written by large
language models (LLMs), and learned surrogates. In earlier work on cardiac pulsed field ablation,
we proposed that solvers return a certificate with their answers. The certificate
consists of an admissible potential and an exactly charge-conserving current, and a small
independent checker turns it into guaranteed error bounds against a written
specification of the physics. Here, the same checker verifies the outputs of other
solvers on the same problem. Specifically, FEniCSx
computes both fields and receives the same bounds as our own solver. Five plausible
bugs in a correct solver, one of them about 0.2\% from the true impedance, all
converge and pass a grid convergence study, yet the certificate
rejects or refutes all five. Of three solvers written by LLMs, one is correct and
certified to 0.1\%, and one never solves the equation but converges. Neural
networks that predict the coefficients of a reduced basis emit admissible fields and
receive valid per-prediction bounds, at most 0.0021\% wide inside their training range
and 0.3--25\% wide outside it, where 90\% conformal intervals cover none of 14 tests.
A Fourier neural operator with free nodal outputs on the certificate grid receives
valid but uninformative bounds, since cells as thin as 1~nm turn small nodal errors
into large energy errors. When it outputs its fields on a triangle mesh refined only
at the electrode corners and is trained in the energy norm with twice as many Fourier
modes, its bounds are 0.57--0.84\% wide in all 12 tests inside its training range
and valid but wide outside it. The guarantee
concerns the stated model and holds for any output, wherever its input lies. It is
informative only when the output is accurate in energy on a grid that resolves the
problem.
\end{abstract}


\section{Introduction}

Simulations governed by partial differential equations (PDEs) used to be set up,
meshed, and checked by an expert, run by run. Increasingly, no expert reviews each run. Design
loops run thousands of simulations unattended, engineers ask large language models
(LLMs) to write solvers, and learned surrogates predict full fields in milliseconds, for instance in cardiac mechanics
\cite{carrara2026}.
None of these settings provides evidence that a particular output is accurate.

The standard forms of evidence do not cover these cases. Code verification, for instance
with manufactured solutions, shows that a code solves its equations correctly on
test problems, but it says nothing about a particular output. A mesh convergence
study (mesh study for short), summarized by the grid convergence index (GCI)
\cite{roache1998,celik2008}, estimates the discretization error of a representative
case and needs judgment about whether the trend is in the asymptotic range. It checks
that the solution converges, but a solver of the wrong equations converges just as
well. A learned surrogate has no grid to refine, so a mesh study does not apply. Test-set and split-conformal
guarantees \cite{angelopoulos2023} are marginal, holding on
average over random draws from the calibration distribution rather than for a given
input or under a distribution shift. Optimal-transport dropout \cite{lorenzon2026} and
other learned predictive distributions are assessed for calibration in the same
averaged sense. Distribution-free conditional coverage is
impossible in general \cite{vovk2013,barber2021}.

In earlier work \cite{dovelos2026}, we separated the question ``was the stated model
solved correctly?'' from the solver that answers it. The solver returns its
answer with a certificate, which consists of a potential that takes the prescribed
electrode and ground values and a current that conserves charge exactly. A small
independent checker verifies the certificate against a written
specification of the physics. Using the Prager--Synge hypercircle
\cite{pragersynge1947} evaluated in ball arithmetic \cite{arb}, it turns the
certificate into guaranteed bounds. These bounds are of the classical
equilibrated-flux type \cite{ladeveze1983,braess2009,ernvohralik2015,repin2008}, and
the contribution of that work was a checker that evaluates them rigorously for a
given output, independently of the solver. Its experiments relied mainly on our own
solver.

This paper applies the unchanged checker to solvers other than our own, against the
same specification and without a reference solution. In particular, it provides:
\begin{itemize}
\item Certificates from FEniCSx \cite{baratta2023}, a widely used finite-element
package. On the certificate grid, its native certificate receives the same bounds as
our own solver, and on its own meshes, its certificates are compared with a mesh
study (Section~\ref{sec:e1}).
\item Five plausible bugs in a correct solver and three solvers written by LLMs. A
mesh study accepts the six wrong programs that ran, while the certificate rejects,
refutes, or leaves unsupported each of them (Section~\ref{sec:e2}).
\item Learned reduced-order surrogates with one and two inputs, which are neural
networks that predict the coefficients of a proper orthogonal decomposition (POD)
basis. They emit exactly admissible certificates and receive valid bounds for each
prediction, inside and outside their training range, and they are compared with
conformal intervals and a functional majorant on the same predictions
(Section~\ref{sec:e3}).
\item Fourier neural operators (FNOs). The bounds of an FNO with output on a uniform
grid are limited by the transfer onto the certificate grid. On the certificate grid
itself, cells as thin as \SI{1}{nm} turn the small nodal errors of an FNO with free
nodal outputs into large energy errors, so its bounds are valid but uninformative.
When its output is mapped onto a triangle mesh with small cells only at the electrode
corners and it is trained in the energy norm with twice as many Fourier modes, an FNO
receives bounds narrower than our 1\% target for a surrogate in all 12 tests inside
its training range (Section~\ref{sec:e3}).
\end{itemize}
Section~\ref{sec:related} reviews related work, and Section~\ref{sec:recap}
summarizes the certificate and the checker, and it defines the test problem and the
possible outcomes. The test problem is linear, steady, elliptic, and axisymmetric, and
Section~\ref{sec:discussion} discusses what carries over to other problems.

\section{Related Work}\label{sec:related}

Computer-assisted proofs evaluate hypercircle bounds in interval arithmetic
\cite{liuoishi2013,liu2021stokes,nakao2019}, and equilibrated fluxes certify
existence and outputs for semilinear problems by post-processing a conventional
solution in floating point \cite{ishizaka2026}. Certified
reduced-basis methods bound outputs over a parameter domain with error estimators
built into the reduced model \cite{rozza2008}. Concurrent work bounds the lesion
volume of a cardiac pulsed-field-ablation catheter with field-dependent conductivity
\cite{talleb2026}. In each case, the bound is computed within the pipeline that
produced the solution. Our checker is a separate program that reads
only a specification and a certificate, as in certifying algorithms
\cite{mcconnell2011} and in small verified checkers of floating-point error
certificates \cite{becker2018}.

For neural PDE solvers, a priori results bound the size of an FNO needed for a given
accuracy \cite{kovachki2021} and its error in reduced floating-point precision
\cite{tu2024}, but they do not bound the error of a given prediction, which is what a
posteriori bounds do. Networks and neural operators can be trained to output additional fields
from which a cheap post-processing step computes a functional error majorant, an
upper bound on the error \cite{fanaskov2024,fanaskov2026}. In a majorant, the flux need
not be divergence-free, and its residual enters with a Friedrichs-type constant
\cite{repin2008}. Rigorous upper bounds also exist for physics-informed networks,
derived from properties of the governing dynamics \cite{hillebrecht2025}. Lower and
upper bounds hold for any approximation whose variational residual can be evaluated
\cite{ernst2025}. Quadrature bounds on network norms, combined with energy
estimates, give verification algorithms that are independent of training
\cite{haugen2025}, and outward-rounded Taylor-model enclosures make a majorant
rigorous for networks that solve obstacle problems \cite{damircheli2026}. Interval
enclosures on adaptively refined boxes give guaranteed function-space norms of
networks \cite{grundler2026}, and interval arithmetic turns a posteriori bounds for
networks that solve ordinary differential equations into machine-checkable proofs
\cite{tanaka2026}. For neural operators, bound propagation in function space
certifies residuals and error bounds over the whole domain \cite{hsiang2026}. That
approach needs the network's weights and bounds what the network can output, whereas
the certificate here needs only the output and bounds its error against the
specification. Closest to the present certificates are networks whose fields are
admissible by construction. A physics-informed network that outputs an admissible
displacement--stress pair gives an upper bound in elasticity \cite{guo2022}, and
networks with an exactly divergence-free velocity and an exactly equilibrated stress
give a two-sided energy enclosure for power-law flow \cite{bukhari2026}. Both concern
a single physics-informed network, and the second states that its bound relies on
converged quadrature rather than an interval enclosure.

LLMs can generate PDE solvers that perform well on representative problems
\cite{codepde2025}, but code that runs without error may still solve the wrong
equations. On a multiphysics benchmark, 39--40\% of generated input files did so,
which was found by reconstructing the encoded PDE and comparing it with the intended
one \cite{song2026}. Such checks are structural and do not bound the error
of an output. Teams of LLM agents also compose preconditioned solvers from Krylov
methods, preconditioners, and neural operators, and rank them by their error against a
direct reference solution \cite{lee2026}. In this paper, one checker verifies a finite-element package,
LLM-written programs, plausible bugs, and learned surrogates without change. It is
independent of every solver, and its bounds include rounding error. Because the learned
reduced-order surrogates supply exactly admissible certificates, no Friedrichs-type
constant enters.

\section{Method and Test Problem}\label{sec:recap}

We first summarize the method of \cite{dovelos2026}. A specification states the domain, the
conductivities $\sigma$, and the boundary conditions of
$\nabla\cdot(\sigma\nabla u)=0$ for the potential $u$. A certificate consists of a
potential $v$ and a current $F$. The potential is piecewise polynomial on a tensor
grid, which we call the certificate grid, and takes the prescribed values on the electrodes and the ground. The current
is the curl of a piecewise-polynomial stream function $\psi$, so it is exactly
divergence-free and has no flux through insulating walls. With
$\lVert X\rVert^2=\int|X|^2\sigma^{-1}\,\mathrm{d}\mu$ and $\mathrm{d}\mu=2\pi r\,\mathrm{d}r\,\mathrm{d}z$, the hypercircle theorem
places $\sigma\nabla u$ on the sphere with center $c=(\sigma\nabla v+F)/2$ and
radius $\eta/2$, where $\eta=\lVert\sigma\nabla v-F\rVert$. This theorem, together
with the Dirichlet and Thomson principles, gives two-sided bounds on the conductance
and on the root-mean-square field of every lesion-map tile, a ring of tissue
\SI{0.25}{mm} wide that is proven lethal, proven spared, or undecided as its interval
lies above, below, or across the lethal threshold (\SI{700}{V/cm} at \SI{1000}{V}
here). The energy of $v$ bounds
the impedance from below and that of $F$ from above.

The checker makes two tests. The specification check verifies admissibility
exactly, by comparing floating-point values for equality. Numerical verification
encloses every cell integral in 256-bit ball arithmetic and bounds the numerical
error. For any admissible certificate, the resulting interval contains the true
value of the stated model. A wrong output therefore fails the specification check,
or the impedance that the run reports, which we call its claim, falls outside the
interval, or
the interval is too wide to support the claim. The experiments with FEniCSx,
plausible bugs, and LLM-written solvers use this 256-bit checker. Most surrogate
experiments involve many predictions and use the vectorized checker of
\cite{dovelos2026} instead, which reproduces its bounds in double-precision ball
arithmetic at 4--15\% of the time our solver needs for both fields.

A solver that returns only a potential is turned into a certificate by an
\emph{adapter}, which fits the stream function to $\sigma\nabla v$ by least squares.
The adapter is untrusted like any solver, and the current it supplies enters the bound. The fit
is a projection onto divergence-free currents, in effect a dual solve, and it returns
a nearly optimal current from any admissible potential. Even
a potential equal to zero away from the boundary receives an upper bound equal to the
true impedance to four decimals (Section~\ref{sec:fno-uniform}). For a
potential-only solver, the upper bound therefore comes from the adapter, and the
certificate tests the solver through the energy of its potential, which gives the
lower bound, and through $\eta$. The adapter costs about as much as a primal solve.

\begin{figure}[t]
\centering
\begin{tikzpicture}[font=\small, node distance=5mm and 7mm,
  box/.style={draw, rounded corners=2pt, align=flush center, inner sep=3pt, minimum height=10mm},
  untrusted/.style={box, fill=orange!10},
  trusted/.style={box, fill=blue!8, thick},
  test/.style={box, fill=blue!8, text width=45mm},
  arr/.style={-{Stealth[length=2mm]}, thick}]
\node[untrusted, text width=50mm] (native) {\textbf{Native:} $v$ and $F$\\
  {\scriptsize FEniCSx with its dual solve; reduced-order surrogates; a neural operator on the certificate grid}};
\node[untrusted, text width=50mm, below=of native] (pot) {\textbf{Potential only:} $v$\\
  {\scriptsize programs written by LLMs; plausible bugs}};
\node[untrusted, text width=50mm, below=of pot] (other) {\textbf{Another grid:} $v$, or $v$ and $F$\\
  {\scriptsize FEniCSx on its own mesh; neural operators on a uniform grid and on a triangle mesh}};
\node[untrusted, dashed, text width=21mm, right=of pot] (ad) {\textbf{Adapter}\\ {\scriptsize fits $F$: a dual solve}};
\node[box, text width=17mm, right=of ad |- native] (cert) {Certificate\\ $(v,F)$};
\node[test, right=9mm of cert] (t1) {\textbf{1.~Specification check}\\
  {\scriptsize boundary values tested exactly}};
\node[test, below=3mm of t1] (t2) {\textbf{2.~Numerical verification}\\
  {\scriptsize hypercircle bounds in ball arithmetic}};
\node[font=\small\bfseries, above=1mm of t1] (chklabel) {Checker};
\node[draw, dashed, thick, rounded corners=3pt, inner sep=2mm, fit=(chklabel)(t1)(t2)] (chk) {};
\node[trusted, text width=41mm, above=4mm of chk] (spec) {\textbf{Specification}\\
  {\scriptsize geometry, conductivities, boundary values}};
\node[box, text width=41mm, below=4mm of chk, fill=green!8] (out) {\textbf{Outcome}\\
  {\scriptsize rejected if test 1 fails; otherwise proven bounds, against which a claim is certified, refuted, or unsupported}};
\draw[arr] (native) -- (cert);
\draw[arr] (pot) -- (ad);
\draw[arr] (other.east) -| node[pos=0.25, above, font=\scriptsize] {transfer, $v$ only} (ad.south);
\draw[arr] (ad.east) -| (cert.south);
\draw[arr] ([yshift=-2.5mm]other.east) -| node[pos=0.3, below, font=\scriptsize] {transfer, with $F$} ([xshift=5mm]cert.south);
\draw[arr] (cert.east) -- (cert.east -| chk.west);
\draw[arr] (spec) -- (chk);
\draw[arr] (chk) -- (out);
\end{tikzpicture}
\caption{Three routes from a solver's output to the checker, with the examples of
this paper. Untrusted components (any solver, the adapter) are orange, every transfer
is untrusted as well, and trusted components are blue. A native certificate supplies both fields, a
potential-only output is completed by the adapter, which is itself a dual solve, and
an output on another grid is transferred first and completed by the adapter if it
has no current. The checker is that of
\cite{dovelos2026}. A certificate that fails the specification check is rejected.
Otherwise, numerical verification returns proven bounds for the stated model, and
claims are judged against them.}
\label{fig:process}
\end{figure}
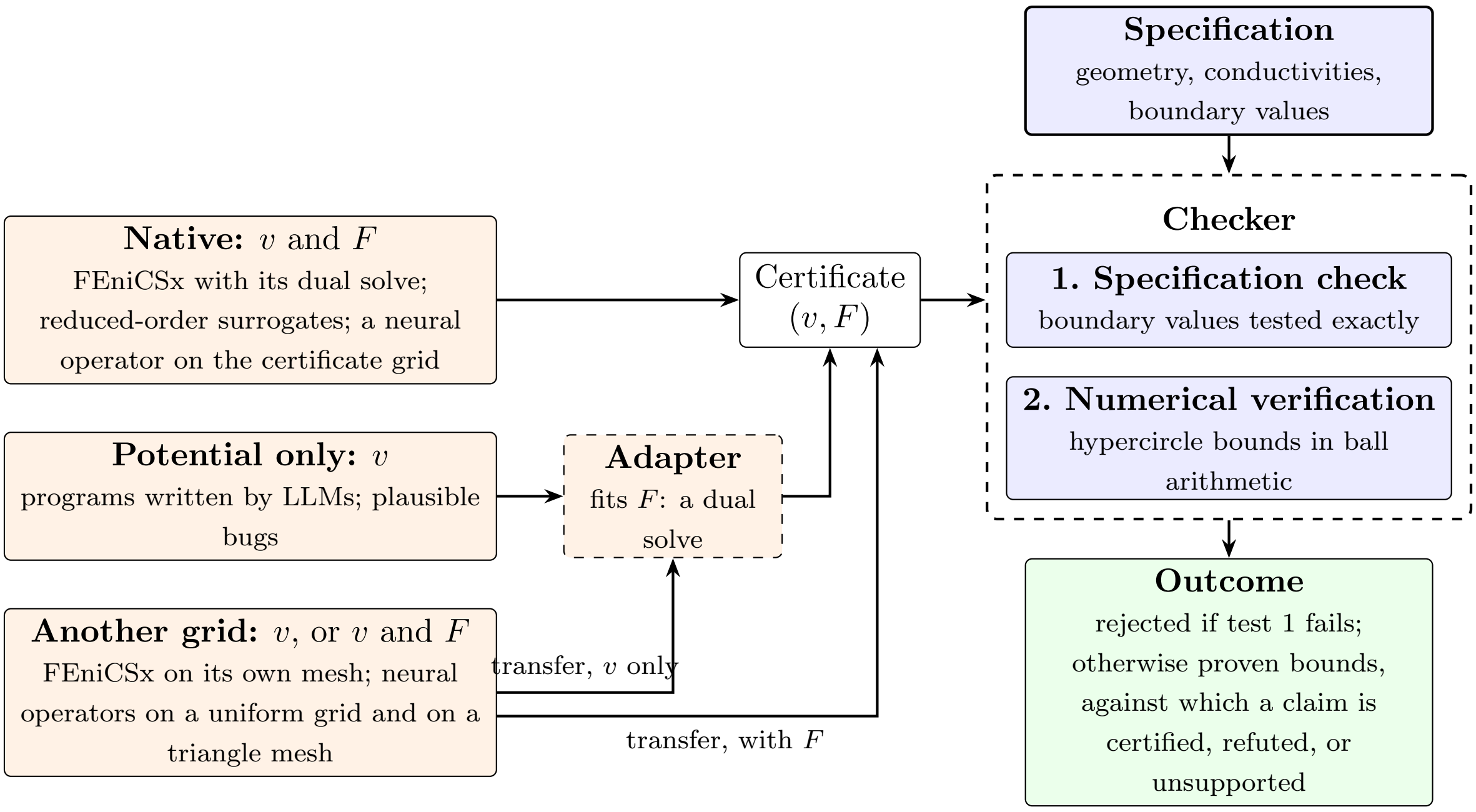

Figure~\ref{fig:process} shows the three ways in which an output reaches the
checker. A solver may return both fields, which we call a \emph{native} certificate.
It may return only a potential on the certificate grid, which the adapter completes.
It may also return a field on another grid, which must first be transferred onto the
certificate grid and given exact boundary values. We call this route the
\emph{transfer} route. If that output includes the stream function, the adapter is not needed. In
every case, the checker returns a guaranteed interval for the stated model. The
routes differ in which bounds come from the solver and in cost, and on the transfer
route, the interval concerns the transferred fields.

All experiments use the monopolar lesion-map problem of \cite{dovelos2026}, in
which a metal electrode of radius \SI{1.25}{mm} and length \SI{3}{mm} is in contact
with myocardium ($\sigma_t=\SI{0.25}{S/m}$) under blood ($\sigma_b=\SI{0.7}{S/m}$, as
stored in binary64), with a grounded box \SI{50}{mm} away. Its certified impedance $Z$,
\SIrange{93.9502742}{93.9502758}{\ohm} (degree 3, $169\times332$ cells, bounds
rounded outward), serves as the reference. Our solver's certificate has
$\eta=1.3\times10^{-5}$ against $\lVert c\rVert\approx0.10$ at \SI{1}{V}, a relative
error of about $10^{-4}$ in the energy norm, and a larger $\eta$ means a less accurate field.
For small errors, the relative width of the impedance interval is about
$(\eta/\lVert c\rVert)^2$, here $2\times10^{-8}$. For each run, the interval is computed
from that run's fields alone, and its claimed impedance is then compared with it. A
run has one of five outcomes. It has \emph{crashed} if it produces no valid output,
and it is \emph{rejected} if its certificate is inadmissible. It is \emph{refuted} if
its claim lies outside the interval by more than $10^{-6}$ relative, a tolerance for
its own rounding, and \emph{unsupported} if the claim lies inside but the interval
does not prove it to 0.1\%. Otherwise, it is \emph{certified}. For a surrogate, we
use a width of 1\% as the target, which is ten times the tolerance for a certified
claim. We call an interval uninformative when it is about as wide as the impedance
or wider.

\section{The FEniCSx Finite-Element Package}\label{sec:e1}

FEniCSx 0.11 \cite{baratta2023} runs in its own environment and exchanges only
files with the checker. The checker sends the specification, and FEniCSx returns a
nodal potential, its own impedance, and, where stated, its stream function. On the certificate
grid, with the same degree-3 tensor elements (410\,248 unknowns on the active cells),
FEniCSx assembles and solves for the potential in \SI{1.9}{s}. Unlike in
\cite{dovelos2026}, where the adapter supplied the current, FEniCSx here also solves
the dual problem. In another \SI{1.9}{s}, it minimizes the complementary energy over
stream functions on a tensor mesh in the coordinates $(r^2,z)$, in which the stream
function is polynomial, with the stream function set to zero on the axis and
constant on the shaft. This certificate, assembled and solved entirely by FEniCSx
from our dual formulation, receives \SIrange{93.950274296}{93.950275789}{\ohm} with
$\eta=1.3\times10^{-5}$, identical to the bound for our own solver. Its ends are the
energies of FEniCSx's potential and current, and as in \cite{dovelos2026}, the two potentials differ by at most \SI{7e-9}{V}.

On its own mesh, FEniCSx provides an independent discretization. Gmsh
\cite{geuzaine2009} meshes the geometry with triangles that are graded geometrically
toward both electrode corners. FEniCSx solves for the potential with Lagrange
elements of degree $k$, and its solution is evaluated at the nodes of the
certificate grid, with the current supplied by the adapter (Table~\ref{tab:e1own}).
With $k=3$ and grading to \SI{1}{nm}, $\eta$ is
$2.0\times10^{-5}$, against $1.3\times10^{-5}$ on the certificate grid, from seven
times fewer unknowns, although the adapter still works on the full certificate grid.
Since the upper bound comes from the adapter, the informative parts are the lower
bound and $\eta$. Quadratic elements give an $\eta$ ten times larger and mostly
undecided tiles, and grading only to \SI{1}{\micro m} leaves the bound valid but
loose. The certificate concerns the potential as evaluated on the certificate grid,
so it includes the transfer error between the two discretizations.

\begin{figure}[tb]
\centering
\includegraphics[width=\textwidth]{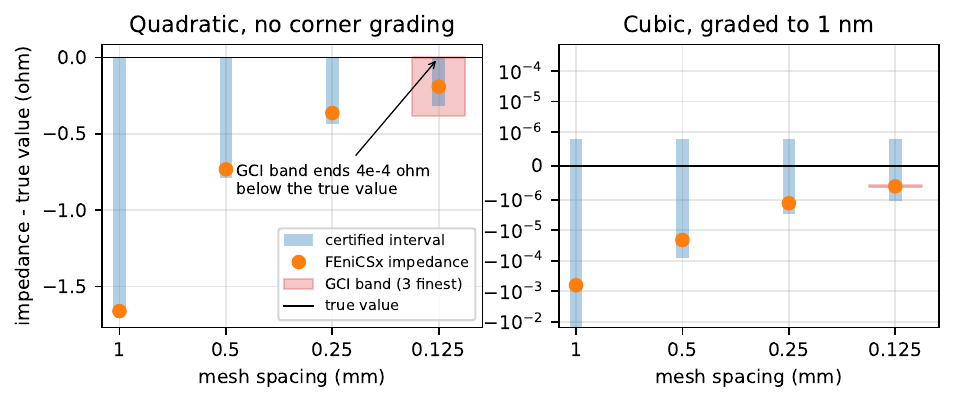}
\caption{FEniCSx's impedance on two mesh families, the certified interval on each
mesh, and the GCI band from the three finest meshes, relative to the true value.
Left: quadratic elements without corner grading, where the band ends just below the
true value. Right (symmetric logarithmic scale): cubic elements graded to
\SI{1}{nm}, where the certified interval contains it and the band, about
\SI{3e-8}{\ohm} below it, lies inside the reference interval.}
\label{fig:gci}
\end{figure}

For a comparison with a mesh study, we refined two families of Gmsh meshes by
factors of two, from 1 to \SI{0.125}{mm}, and scaled the grading slope and the
far-field growth with the spacing. FEniCSx's impedances on the three finest meshes
give the GCI \cite{celik2008}, classified by the convergence ratio $R$
\cite{stern2001}. On cubic elements graded to \SI{1}{nm}, the triplet converges at
order 4.8, and the GCI is consistent with the reference. On quadratic elements
without grading, as a typical user would mesh the problem, the triplet converges
monotonically ($R=0.47$) at an observed order of 1.09. The GCI band with a safety
factor of 1.25, \SIrange{93.5687}{93.9499}{\ohm}, misses the true impedance
narrowly, by \SI{4e-4}{\ohm}. Roache reserves the factor 1.25 for an observed order
that matches the formal one \cite{roache1998}. With his more conservative factor of
3, the band contains the true value, so the GCI succeeds only with this choice of
safety factor. The certified interval contains the true value on every mesh of both
families, as it must, and on the ungraded family, its undecided tiles show that the
mesh is inadequate near the electrode. Figure~\ref{fig:gci} shows both families.
\section{Plausible Bugs and LLM-Written Solvers}\label{sec:e2}
\begin{table}[t]
\centering
\caption{FEniCSx Potential on Its Own Triangular Mesh}\label{tab:e1own}
\small
\resizebox{\textwidth}{!}{%
\begin{tabular}{lrcccc}
\toprule
FEniCSx & Unknowns & Certified impedance (\si{\ohm}) & $\eta$ & Tiles, median & Claim proven\\
\midrule
P2, \SI{0.25}{mm}, \SI{1}{\micro m} grading & 21\,591 & 93.80965--93.95028 & $4.0\times10^{-3}$ & 200\% & no\\
P2, \SI{0.25}{mm}, \SI{1}{nm} grading & 26\,310 & 93.94985--93.95028 & $2.2\times10^{-4}$ & 32\% & yes\\
P3, \SI{0.25}{mm}, \SI{1}{nm} grading & 58\,927 & 93.950272--93.950276 & $2.0\times10^{-5}$ & 3.0\% & yes\\
Certificate grid, native (Q3) & 410\,248 & 93.950274296--93.950275789 & $1.3\times10^{-5}$ & 1.9\% & yes\\
\bottomrule
\end{tabular}}
\par\vspace{2pt}\footnotesize\raggedright Rows 1--3: current from the adapter. P$k$: Lagrange triangles of degree $k$; Q3: degree-3 tensor elements. Tiles: median relative width. ``Claim proven'': FEniCSx's impedance proven to $10^{-5}$. Bounds rounded outward.
\end{table}

Each solver receives the same specification, which consists of the PDE, the
geometry, the boundary conditions, and a file interface. It does not receive the
lists of boundary nodes, so locating them is part of the task. Each solver runs in a
scratch folder at three systematically refined resolutions (1, 0.5, and
\SI{0.25}{mm}) and writes its potential on its own tensor grid. The checker certifies
this potential at degree 1 on that grid, with the current from the adapter, and the
claimed impedances on the three grids form a mesh study.

Earlier work \cite{dovelos2026} showed that a mesh study accepts two wrong-physics
variants of our solver. Here, we applied five plausible one-line bugs, including
those two, to a different correct solver, namely the program written by Claude
Opus~5.5 and described below. Table~\ref{tab:mutants} shows the result. All five
variants converge, so a mesh study accepts all five, but the certificate catches
every one. The hardest case is the top boundary left insulating instead of
grounded. Its impedance is about 0.2\% from the true value, and only the
specification check exposes it. A claim above the true value is refuted by the upper
bound, which here comes from the adapter, and a claim below it is refuted by the
energy of the solver's own potential. In \cite{dovelos2026}, the swapped
conductivities were unsupported rather than refuted, because the solver supplied its
own current, fitted with the wrong conductivities. Here, the adapter fits the
current with the specified ones.

\begin{table}[t]
\centering
\caption{Plausible Bugs in a Correct Solver}\label{tab:mutants}
\small
\resizebox{\textwidth}{!}{%
\begin{tabular}{lccc}
\toprule
One-line bug & Its impedance (\si{\ohm}) & Mesh study & Certificate\\
\midrule
Top boundary insulating instead of grounded & 94.16 (0.2\% off) & converging, $R=0.40$ & rejected\\
Only the electrode's bottom face driven & 390.9 & converging, $R=0.31$ & rejected\\
Tissue and blood conductivities swapped & 136.7 & converging, $R=0.45$ & refuted (88.62--93.98)\\
Axisymmetric weight $r$ missing & 0.393 & converging, $R=0.29$ & refuted (40.34--93.98)\\
Factor $2\pi$ missing in the current & 590.0 & converging, $R=0.40$ & refuted (93.90--93.98)\\
\bottomrule
\end{tabular}}
\end{table}

Of the solvers written by LLMs, two programs came from a local open model
(Qwen~3.8, 27B parameters, through the Ollama interface, seeds 1 and 2, temperature
0.7, reasoning mode off). One came from a frontier model (Claude Opus~5.5, run as a
fresh Claude Code session given the specification and an instruction not to use
tools, with default sampling and no seed). The two setups differ, and three programs
are too few to estimate a rate. Table~\ref{tab:e2} summarizes the results.

\begin{table}[t]
\centering
\caption{Solvers Written by Language Models}\label{tab:e2}
\small
\begin{tabular}{p{0.19\textwidth}p{0.25\textwidth}p{0.3\textwidth}p{0.13\textwidth}}
\toprule
Solver & Mesh study on its claims & Certificate, finest grid & Correct to 0.1\%\\
\midrule
Claude Opus 5.5 (88 lines) & converging ($R=0.40$), band contains the true value & certified: claim \SI{93.8997}{\ohm} proven to 0.1\%, $\eta=3.0\times10^{-3}$ & yes\\
Qwen 3.8, seed 2 (814 lines) & converging ($R=0.44$, order 1.2), band \SIrange{5.5e-5}{7.1e-5}{\ohm} & unsupported: \SIrange{6.3e-5}{93.98}{\ohm}, $\eta=126$ & no (factor $1.5\times10^{6}$)\\
Qwen 3.8, seed 1 (542 lines) & none & crashed: singular system, NaN potential, exit code 0 & no\\
\bottomrule
\end{tabular}
\end{table}

The Qwen program with seed 2 sets the boundary values but never moves them to the
right-hand side, so it solves a homogeneous system. Every free node is at \SI{0}{V},
and the field drops from the electrode value to zero within one cell. Its boundary
values are exactly right, so the potential is admissible, and its claims change
smoothly with the grid, so a mesh study accepts their convergence. The certificate
does not support its claim, and its $\eta$ is four orders of magnitude above that of
the correct solver. An order-of-magnitude check, or code verification with a
manufactured solution, would also have caught this program, but an
order-of-magnitude check would miss four of the five bugs above. The Claude program
is a correct bilinear finite-element code. Its claim lies $6\times10^{-10}$ relative
below the rigorous lower bound, which is consistent with rounding in its current
sum. Among the seven programs that ran, convergence under refinement did not
separate correct solvers from wrong ones, whereas the certificate did, without a
reference solution. For these potential-only programs, its upper bound comes from
the adapter's dual solve.

\section{Learned Surrogates}\label{sec:e3}

The checker accepts an admissible pair from any source, and a surrogate can
supply one through the routes of Figure~\ref{fig:process}. For a native certificate,
the surrogate predicts the free nodal values of both the potential and the stream
function on the certificate grid, and the boundary and wall values are imposed
exactly, so the certificate is entirely the surrogate's. Reduced-order surrogates
with one input, the conductivity ratio $\rho$, and with two, $\rho$ and the electrode
length $L$, follow the native route (Section~\ref{sec:rom}).
A neural operator on a uniform grid follows the transfer route
(Section~\ref{sec:fno-uniform}), and one on the certificate grid the native route
again (Section~\ref{sec:fno-grid}). An operator whose output is mapped onto a
triangle mesh follows the transfer route without the adapter
(Section~\ref{sec:fno-tri}). Section~\ref{sec:baselines} compares the
certificate with conformal intervals and a functional majorant, and
Table~\ref{tab:training} gives the training details.

\begin{table}[t]
\centering
\caption{The Surrogates and Their Training}\label{tab:training}
\footnotesize
\newcolumntype{L}[1]{>{\raggedright\arraybackslash}p{#1}}
\begin{tabular}{L{0.11\textwidth}L{0.17\textwidth}L{0.25\textwidth}L{0.27\textwidth}L{0.07\textwidth}}
\toprule
Surrogate & Training data & Model & Loss and optimizer & Cost\\
\midrule
Reduced order, one input & 17 solutions, $\rho\in[1.7,3.75]$ & POD, 6 modes per field; network $1\to16\to16\to13$, tanh & Squared error of standardized POD coefficients and log impedance; Adam, rate $3\times10^{-3}$, full batch, 20\,000 steps & \SI{1}{s}, CPU\\
\addlinespace
Reduced order, two inputs & 64 solutions, Latin hypercube in $\rho\in[1.7,3.75]$, $L\in[2.5,3.5]$\,mm & POD, 12 modes per field; network $2\to32\to32\to24$, tanh & Squared error of standardized POD coefficients; optimizer as above & \SI{1}{s}, CPU\\
\addlinespace
Neural operator, certificate grid & The same 64 solutions, on the $997\times508$ node grid & FNO, $48\times24$ modes, 32 channels, 4 layers, 2.6 million parameters; inputs $\log\rho$, $L$, node coordinates, mask; outputs $v$ and $\psi$ & Nodal squared error: Adam, batch 2, cosine schedule, 150 epochs on every second node at $2\times10^{-3}$, then 30 on all nodes at $2\times10^{-4}$. Energy of the error (Section~\ref{sec:fno-grid}): 60 and 40 epochs at $10^{-3}$ and $10^{-4}$ & 46 and \SI{38}{min}, GPU\\
\addlinespace
Neural operator, triangle output & 64, 128, or 256 FEniCSx solutions on the triangle mesh, nested sets of Latin hypercube samples in the same box & The FNO above with $48\times24$ or $96\times48$ modes (2.6 or 9.8 million parameters; added modes initialized to zero), output interpolated onto cubic elements on 12\,911 triangles & Relative energy of the error with the stiffness matrices of FEniCSx, from the nodal-loss weights; Adam, rate $3\times10^{-5}$, cosine schedule, batch 2, 4\,480 steps; model chosen on 24 calibration members & about \SI{2}{h}, GPU\\
\addlinespace
Neural operator, uniform grid & 32 solutions, $\rho\in[1.7,3.75]$, on a $256\times512$ grid & FNO, $32\times32$ modes, 32 channels, 4 layers; inputs $\log\rho$, $r$, $z$, mask; output $v$ & Squared error over the domain and the electrode; Adam, rate $2\times10^{-3}$, cosine schedule, batch 4, 400 epochs & \SI{8.4}{min}, GPU\\
\bottomrule
\end{tabular}
\par\vspace{2pt}\footnotesize\raggedright The reduced-order networks are written in NumPy; the neural operators use the neuraloperator library \cite{kossaifi2024} with PyTorch on the graphics processor of a MacBook Pro (M5 Pro, 48 GB). Training solutions come from our solver at degree 3, except for the triangle output,
whose training solutions come from FEniCSx.
\end{table}

\subsection{Reduced-Order Surrogates}\label{sec:rom}
The first surrogate is a neural network on a reduced basis \cite{hesthaven2018}
that maps the conductivity ratio $\rho=\sigma_b/\sigma_t$ to the fields. A POD of
the training solutions gives a basis for the free nodal values of the potential and
of the stream function, and a small network maps $\log\rho$ to the POD coefficients
and to the logarithm of the impedance. The electrode and ground values and the wall
values of the stream function are exact, so every prediction is a certificate
supplied entirely by the surrogate, with no adapter. A prediction takes about a
millisecond and the vectorized check about a second, so the check dominates the
cost.

The second surrogate takes two inputs, $\rho$ and the electrode length
$L\in[2.5,3.5]$\,mm. Changing $L$ moves only the grid lines above the tissue, so every
member's certificate grid is the grid for $L=\SI{3}{mm}$ with those lines
mapped piecewise linearly, and all members share one node numbering. On a mapped grid,
our solver's certificate has the same quality as on the original grid
($\eta=1.3\times10^{-5}$ at $L=\SI{2}{mm}$). Both surrogates are tested inside and
outside their training range (Table~\ref{tab:e3}).

\begin{table}[t]
\centering
\caption{Certifying Reduced-Order Surrogates per Prediction}\label{tab:e3}
\small
\resizebox{\textwidth}{!}{%
\begin{tabular}{llcccc}
\toprule
Surrogate & Tests & True value inside certified interval & Certified width (rel.) & Error of the impedance from its potential & Conformal interval (90\%) covers\\
\midrule
One input & inside (6) & 6 & $2\times10^{-8}$--$2\times10^{-5}$ & $8\times10^{-9}$--$1.2\times10^{-6}$ & 5\\
 & outside (6) & 6 & 0.8--15\% & $2.6\times10^{-4}$--$1.3\times10^{-2}$ & 0\\
Two inputs & inside (12) & 12 & $6.4\times10^{-7}$--$1.9\times10^{-5}$ & $1.1\times10^{-7}$--$2.4\times10^{-6}$ & 12\\
 & outside (8) & 8 & 0.27--25\% & $1.4\times10^{-4}$--$2.8\times10^{-2}$ & 0\\
\bottomrule
\end{tabular}}
\par\vspace{2pt}\footnotesize\raggedright One input: test ratios 1.8--3.6 inside, 0.7--1.3 and 5--10 outside. Two inputs: 12 random members inside; $(\rho, L/\mathrm{mm})$ = (0.7, 3), (1, 3), (5, 3), (10, 3), (2.8, 2), (2.8, 4), (0.7, 2), (10, 4) outside. Conformal intervals: Section~\ref{sec:baselines}.
\end{table}

Every interval contains the true impedance, as the theorem guarantees, which also
serves as a check of the implementation. Inside the range, the impedance is
certified to $2\times10^{-8}$--$2\times10^{-5}$ with one input and to
$6\times10^{-7}$--$2\times10^{-5}$ with two. The tile intervals vary more, with
median widths of 3--139\% for one input, because every tile uses the global $\eta$.
With two inputs, the lesion depth at \SI{1000}{V} and \SI{700}{V/cm} is bracketed on
both sides, by a proven-lethal and a proven-spared tile on the axis, in 6 of 12
in-range tests, and every bracket overlaps that of the
reference. Inside the range, $\eta$ is $2\times10^{-5}$--$9\times10^{-4}$ with one
input and $9\times10^{-5}$--$1.1\times10^{-3}$ with two. The two-input training
fields are stored in single precision, so a stored reference field itself receives
$\eta\approx1.6\times10^{-4}$, and the projection onto 12 modes removes part of this
rounding error. Outside the range, $\eta$ grows by one to three orders of magnitude,
the intervals stay valid, and the tiles become undecided.

\begin{figure}[t]
\centering
\includegraphics[width=0.7\textwidth]{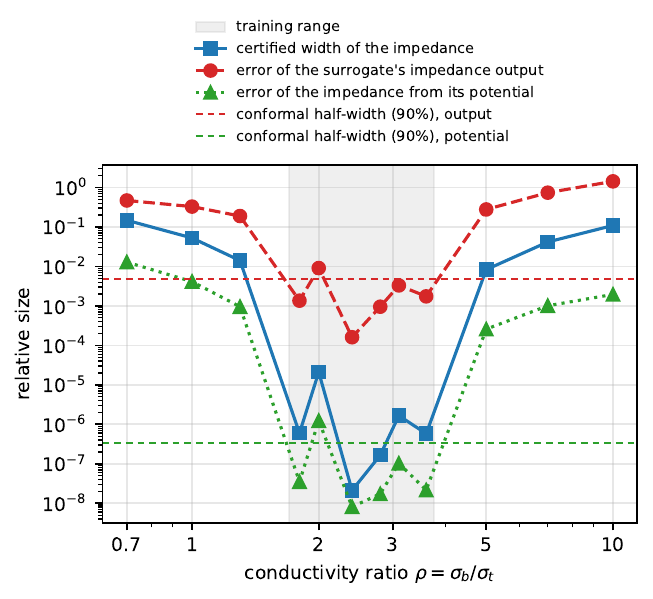}
\caption{The one-input reduced-order surrogate at twelve test ratios. Squares show
the relative width of the certified impedance interval. Circles show the relative
error of the surrogate's own impedance output, and triangles that of the impedance
computed from its potential, both against a reference solve. Dashed lines show the
half-widths of the 90\% conformal intervals around each, calibrated on 20 ratios in
the training range (shaded). Outside this range, both errors exceed their conformal
half-widths, while every certified interval contains the true value.}
\label{fig:surrogate}
\end{figure}

The one-input surrogate also predicts the impedance directly. This output has errors
of 0.02--0.9\% inside the range and 19--140\% outside, and it is refuted at every
test point, by margins far larger than the certified width. The fields of the
surrogate are more accurate than its scalar output, so the impedance is best read
from the certified fields. A first version, which predicted the shaft value of the
stream function separately from its neighbors, produced spurious currents next to
the shaft and uninformative upper bounds, while its lower bounds stayed accurate.
Figure~\ref{fig:surrogate} compares the certified width with both errors over the
twelve test ratios.

\subsection{A Neural Operator on a Uniform Grid}\label{sec:fno-uniform}
An untuned FNO \cite{li2021} was trained on 32 ratios to map $\log\rho$, $r$, $z$,
and a domain mask to the potential on a uniform $256\times512$ grid over the whole
domain, with \SI{0.2}{mm} spacing (Table~\ref{tab:training}). A prediction takes
\SI{16}{ms}, and the final training loss corresponds to about \SI{0.6}{mV}
root-mean-square on a \SI{1}{V} field. The output is not admissible. Each prediction
was therefore interpolated onto the certificate grid, given exact electrode and
ground values, either by overwriting them or by spreading the electrode mismatch
over one FNO grid spacing, and completed with the adapter's current. Two controls
followed the same route, namely the reference solution sampled on the same uniform
grid and a zero field. Table~\ref{tab:fno} summarizes the results.

\begin{table}[t]
\centering
\caption{A Uniform-Grid Output, Certified After the Fact (12 Test Ratios)}\label{tab:fno}
\small
\resizebox{\textwidth}{!}{%
\begin{tabular}{lcccc}
\toprule
Input to the route & $\eta$ & Certified lower bound (\si{\ohm}) & Upper bound & Tiles\\
\midrule
FNO, boundary values overwritten & 8.8--42 & 0.0006--0.013 & = true value & undecided\\
FNO, mismatch spread & 2.7--11 & 0.009--0.14 & = true value & undecided\\
Reference sampled on the uniform grid & 10--38 & 0.0007--0.010 & = true value & undecided\\
Zero field (boundary values only) & 17--65 & 0.0002--0.0034 & = true value & undecided\\
\bottomrule
\end{tabular}}
\par\vspace{2pt}\footnotesize\raggedright True impedances \SIrange{30}{264}{\ohm}; a solve on the certificate grid gives $\eta\approx10^{-5}$.
\end{table}

All intervals contain the true value, and none is informative on the lower side,
which is the side that the transferred potential determines. The upper bound comes from the adapter, and the zero field
receives it too. The reference solution sampled on the same grid does no better than
the FNO, because no sampling on a uniform \SI{0.2}{mm} grid resolves the singular
field at the electrode edge; this control is also deliberately pessimistic, because
we sample the insulating shaft as \SI{0}{V}. The limiting factor is therefore the
transfer. With this
transfer, no field sampled at \SI{0.2}{mm} received an informative bound. Compared
with the reference field sampled on the same uniform grid at $\rho=2.8$, the FNO errs by
\SI{0.15}{mV} on average more than \SI{0.5}{mm} from the electrode, but by
\SI{16}{mV} on average and \SI{64}{mV} at most within \SI{0.5}{mm} of it. The
training loss, dominated by the far field, does not show these errors. Section~\ref{sec:fno-grid} avoids the transfer.

\subsection{A Neural Operator on the Certificate Grid}\label{sec:fno-grid}
Without a transfer, an FNO works on the index grid of the certificate, in which node
$(j,i)$ of the degree-3 node array is one grid point, whatever its physical spacing.
Its spectral convolutions treat this index grid as uniform, although a grid graded
from \SI{1}{nm} to millimetres is far from uniform. Our operator maps $\log\rho$ and
$L$, given as constant fields, together with the node coordinates and the active
mask, to both the potential and the stream function, and it uses the same 64
training members as the two-input surrogate. The boundary values are imposed exactly
after the forward pass, so each prediction is a certificate, again without an
adapter. A prediction takes \SI{0.27}{s}.

\begin{figure}[t]
\centering
\includegraphics[width=0.9\textwidth]{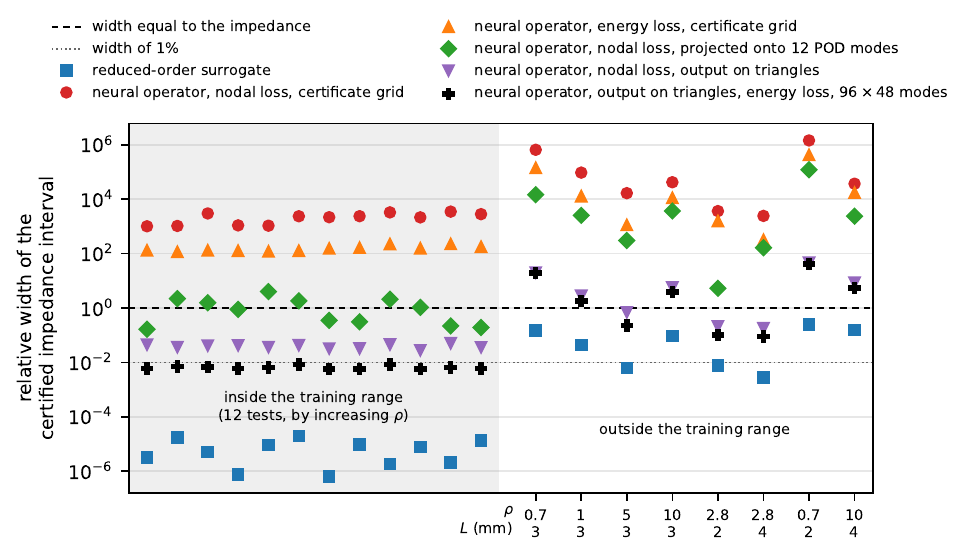}
\caption{Relative width of the certified impedance interval for each of the 20
tests of the two-input surrogates (shaded: inside the training range). Every interval contains the true impedance. The reduced-order
surrogate receives widths of $6\times10^{-7}$--$2\times10^{-5}$ inside the range and 0.3--25\%
outside it. Inside the range, the neural operator's intervals are $10^2$ to $3.5\times10^3$
times the impedance after training with either loss, and 0.16--4 times after
projection onto the 12 POD modes of the reduced-order surrogate. With its output
mapped onto a triangle mesh, the operator trained with the nodal loss receives widths
of 2.8--5.1\%, and after training in the energy norm with $96\times48$ modes,
0.57--0.84\% (Section~\ref{sec:fno-tri}).}
\label{fig:operator}
\end{figure}

When trained with a nodal squared error, the operator has small nodal errors,
\SI{0.36}{mV} at the median and \SI{9}{mV} at most on an in-range test.
Nevertheless, every certified interval, valid in all 20 tests, is wider than the
impedance itself by a factor of $10^3$ or more, and the impedance computed from its
potential is 98--99\% too low (Table~\ref{tab:fno-grid} and
Figure~\ref{fig:operator}). The reason lies in the certificate grid. It is a tensor
grid graded to \SI{1}{nm} at the electrode corners, so the graded lines cross the
whole domain, and more than 9\,000 active cells are thinner than \SI{10}{nm}, with
aspect ratios up to $3\times10^6$. Across most of these cells, the true potential
changes by less than \SI{1}{\micro V}, so a nodal error of \SI{0.1}{mV} that differs
between the two sides of a cell puts a spurious field of $10^5$\,V/m into it. The
fields of the solver, and any combination of POD modes, are consistent across these
cells, but the outputs of a network that predicts each nodal value freely need not
be.

Most of the width of the nodal-loss operator's intervals comes from the cells next to
the electrode and the shaft, where exactly imposed boundary values meet cells
\SI{1}{nm} thick along the whole surface. In four in-range tests, we replaced the
operator's values within \SI{10}{\micro m} of the electrode and the shaft by those of
the reference solution, which reduces the width from $10^3$--$3.5\times10^3$ to 5--17
times the impedance. The same replacement within \SI{10}{\micro m} of every graded
grid line reduces it to 1.0--1.6 times. A certificate grid graded less finely has
fewer thin cells but resolves the lesion map less sharply. Graded to
\SI{1}{\micro m} instead of \SI{1}{nm}, it has no cell thinner than
\SI{1}{\micro m}. On that grid, our solver certifies the impedance to a relative
width of $1.1$--$1.3\times10^{-5}$ but brackets the lesion depth in only 6 of the 12
tests, against all 12 on the grid graded to \SI{1}{nm}. We resampled the operator's error
onto this coarser grid and added it to the solver's fields, which still gives widths
of 0.26--0.58 times the impedance.

We then fine-tuned the operator with a loss that measures what the certificate
measures, namely the energy of the error of the potential and the complementary
energy of the error of the stream function, with the boundary values imposed inside
the network's output. This lowers the energies of the errors, relative to the
energies of the fields and averaged over the training fields, from 78 and 1\,700 for the potential and the stream function to 45
and 27, and it lowers $\eta$ four- to sixfold. Both errors remain much larger than
the energies of the fields themselves, so the fine-tuned operator is still far from
accurate, and its bounds remain uninformative.

A simple post-processing step projects each prediction, with its boundary values
imposed, onto the
first 12 POD modes of the training solutions, which form the basis of the
reduced-order surrogate, by least squares on the free nodal values. The projected
fields are consistent across the thin cells. For the operator trained with the nodal
loss, the certified width falls from $10^3$--$3.5\times10^3$ to 0.16--4 times the
impedance, and $\eta$ falls to 0.04--0.18. This is still short of an informative
bound, although the same basis certifies the projected reference fields to
$3\times10^{-7}$--$3\times10^{-6}$ (Table~\ref{tab:fno-grid}). The POD coefficients
explain the difference. On an in-range test, the reference coefficients fall below
$10^{-3}$ after the fourth mode, whereas the operator's errors put about $10^{-2}$
into every mode, in directions that the solution barely uses but that carry energy.
More modes, 24 or 48, widen the intervals, and for the operator fine-tuned in the
energy norm, the projection widens them as well ($\eta=15$--20). The reduced-order
surrogate avoids this problem because its network predicts the coefficients
themselves, each scaled by its own spread. The projection minimizes the nodal
$\ell^2$ norm, whereas the certificate measures the energy norm. With the basis available, a 12-mode
Galerkin solve would give a certificate without any network, so the projection tests
only whether the operator's own output can be corrected cheaply.

On the certificate grid itself, a local correction removes most of the effect of the
thin cells. Neural operators have been combined with relaxation methods inside
iterative solvers before \cite{zhang2024}, and here the sweeps follow the operator's
prediction. Started from the nodal-loss operator, five symmetric line Gauss--Seidel
sweeps of our solver's two linear systems give widths of 0.12--0.31\% inside the
range, and 0.094--0.21\% after the operator is fine-tuned through the sweeps. Each
sweep solves for all nodes of a grid line at once, alternately along $r$ and $z$. The
prediction and the sweeps take \SI{1.0}{s}, and assembling the two systems takes
another \SI{1.25}{s}. The same sweeps started from zero give no informative bound,
and outside the range they leave every width above 1\%. The conjugate gradient
method with a two-level p-multigrid preconditioner reaches a relative width of
$1.4\times10^{-5}$ from zero in three iterations and \SI{1.2}{s}, again without
assembly, and the operator's prediction as a starting point saves one iteration of
\SI{0.18}{s}. These corrections iterate on our solver's linear systems, whereas
Section~\ref{sec:fno-tri} changes only the operator's output.

\begin{table}[t]
\centering
\caption{A Neural Operator on the Certificate Grid (12 Tests Inside and 8 Outside the Training Range)}\label{tab:fno-grid}
\footnotesize\setlength{\tabcolsep}{3pt}
\begin{tabular}{lcccc}
\toprule
Surrogate & $\eta$, inside & Width, inside & $Z$ from $v$, inside & Contains $Z$\\
\midrule
Operator, nodal loss & 3.0--6.8 & $10^3$--$3.5\times10^3$ & 0.98--0.99 & 20 of 20\\
Operator, energy loss & 0.78--1.1 & $1.1\times10^2$--$2.2\times10^2$ & 0.98 & 20 of 20\\
Operator, projected & 0.037--0.18 & 0.16--4.0 & -- & 20 of 20\\
Reduced order & $8.8\times10^{-5}$--$1.1\times10^{-3}$ & $6.4\times10^{-7}$--$1.9\times10^{-5}$ & $1.1\times10^{-7}$--$2.4\times10^{-6}$ & 20 of 20\\
\bottomrule
\end{tabular}
\par\vspace{2pt}\footnotesize\raggedright Width: relative width of the certified impedance interval; $Z$ from $v$: relative error of the impedance from the energy of the predicted potential; energy loss: after fine-tuning in the energy norm; projected: the operator trained with the nodal loss, projected onto the 12 POD modes of the reduced-order surrogate. Same training members, tests, and grid for all; a solve on this grid gives $\eta\approx10^{-5}$.
\end{table}

\subsection{A Neural Operator With Output on a Triangle Mesh}\label{sec:fno-tri}
Because the thin cells of the certificate grid cause most of the width, we kept the
operator and first changed only its output space. A fixed sparse map interpolates the
output bilinearly between neighboring nodes of the index grid, in $(r,z)$ for the
potential and in $(r^2,z)$ for the stream function, and it maps both fields onto
cubic Lagrange elements on a conforming mesh of 12\,911 triangles, graded by Gmsh
toward the electrode corners as in Section~\ref{sec:e1}. Its smallest edge is
\SI{0.94}{nm}, its largest aspect ratio is 1.8 (1 for an equilateral triangle), and
every triangle whose longest edge is shorter than \SI{1}{\micro m} lies within
\SI{5}{\micro m} of a corner. As $L$ changes, each vertex moves with the
piecewise-linear map that moves the lines of the certificate grid
(Section~\ref{sec:rom}). An added line at $z=\SI{3}{mm}$, where this map changes
slope, makes it affine on every triangle, so the interpolation is the same for every
member. The boundary values are then imposed exactly, and both fields are evaluated
at the nodes of the certificate grid and checked there, as for FEniCSx on its own mesh in
Section~\ref{sec:e1}, but without the adapter, since the output includes the stream
function. The certificate therefore bounds the cubic interpolant of the triangle
fields on the certificate grid, with the boundary values imposed again there. For the
solutions of FEniCSx, this interval is 8--14\% narrower than a floating-point
evaluation on the triangle fields themselves. FEniCSx solves the primal and dual problems on this mesh in
\SI{0.7}{s}. Through this route, its solutions receive $\eta=1.4$--$1.8\times10^{-5}$,
and they also serve as the training data.

\begin{table}[t]
\centering
\caption{A Neural Operator With Output on a Triangle Mesh (12 Tests Inside and 8 Outside the Training Range)}\label{tab:fno-tri}
\footnotesize\setlength{\tabcolsep}{4pt}
\begin{tabular}{lccccc}
\toprule
Training & Members & Modes & Below 1\%, inside & Width, inside & Width, outside\\
\midrule
Nodal loss, no retraining & 64 & $48\times24$ & 0 of 12 & 2.8--5.1\% & 0.18--46\\
Energy loss on triangles & 64 & $48\times24$ & 5 of 12 & 0.89--1.37\% & 0.079--40\\
Energy loss on triangles & 128 & $48\times24$ & 5 of 12 & 0.92--1.32\% & 0.098--40\\
Energy loss on triangles & 256 & $48\times24$ & 5 of 12 & 0.93--1.30\% & 0.098--41\\
Energy loss on triangles & 256 & $96\times48$ & 12 of 12 & 0.57--0.84\% & 0.089--42\\
\bottomrule
\end{tabular}
\par\vspace{2pt}\footnotesize\raggedright Width: certified width of the impedance interval relative to the impedance, through the certificate grid; below 1\%: the target for a surrogate; outside the range, in multiples of the impedance. All 100 intervals contain the true impedance. A prediction takes \SI{0.27}{s} for every variant.
\end{table}

Mapped onto the triangles without retraining, the predictions of the nodal-loss
operator receive bounds 2.8--5.1\% wide inside the range, instead of
$10^3$--$3.5\times10^3$ times the impedance. We then trained the operator through the
fixed map, starting from the nodal-loss weights. The loss is the energy of the
difference between its fields and the primal and dual solutions of FEniCSx, relative
to the energy of those solutions and computed with the stiffness matrices of the
same mesh. Because the output and these solutions lie in the same finite-element
spaces and take the same boundary values, Galerkin orthogonality makes this loss
equal to $(\eta/\lVert c\rVert)^2$ of the triangle fields less that of the reference
solutions, about $2\times10^{-8}$, apart from the quadrature of FEniCSx. Inside the
range, the certified width therefore follows the loss, with a slope of about 1.04 on
logarithmic axes. For each variant, we kept the weights with the lowest mean loss on
the 24 calibration members (Table~\ref{tab:training}). With the original $48\times24$
Fourier modes, training narrows the intervals to 0.89--1.37\%, below the target in 5
of the 12 tests, and training sets of 128 and 256 members leave them almost unchanged,
at 0.92--1.32\% and 0.93--1.30\%. Twice as many modes in each direction, $96\times48$ with
9.8 million parameters, narrow them to 0.57--0.84\%, below the target in all 12 tests
(Table~\ref{tab:fno-tri} and Figure~\ref{fig:operator}). Inside the training range,
more Fourier modes therefore narrowed the intervals, whereas more training members
did not. A prediction takes \SI{0.27}{s} and its check about \SI{1.1}{s}. FEniCSx
solves the problem on the same mesh in \SI{0.7}{s}, and the same check certifies its
solution to a relative width of $3\times10^{-8}$. On this two-dimensional problem, the
certified prediction therefore saves about \SI{0.4}{s} of \SI{1.8}{s}, but its
interval is 2--3$\times10^5$ times wider.

Outside the training range, no variant receives bounds narrower than the target, and
every interval contains the true impedance. Mapping the output onto the triangles
narrows the widths of the nodal-loss operator from $2.4\times10^3$--$1.4\times10^6$ to
0.18--46 times the impedance. Training in the energy norm narrows them by a factor of
1.1--2.5, whereas more training members and more Fourier modes change them little, so
for the larger operator, they are 0.089--42 times the impedance. For the operator
with $48\times24$ modes, about two thirds of the energy of the potential's error lies
more than \SI{1}{cm} from the electrode corners, against a fifth inside the range, so
its error outside the range is global.

\subsection{Conformal Intervals and a Functional Majorant}\label{sec:baselines}
Split conformal prediction \cite{angelopoulos2023} calibrates an interval around a
point estimate on held-out members of the training distribution. Here, it uses 90\%
nominal coverage, with 20 members for the one-input surrogate and 24 for the
two-input one. Inside the range, it covers 5 of 6 and 12 of 12 tests. Around the
impedance from the potential, its intervals are $6.6\times10^{-7}$ and
$3.2\times10^{-5}$ of the impedance wide, comparable with the certified widths. Outside
the range, it covers none of the 14 tests, because the errors there exceed the
calibrated half-width by a factor of 9 or more, while every certified interval
contains the true value (Table~\ref{tab:e3} and Figure~\ref{fig:surrogate}). The neural
operators on the certificate grid, trained with the nodal and the energy loss,
estimate the impedance 98\% too low. Their conformal intervals inside the range are
1.8--3.6 times the impedance wide, narrower than their certified ones, and cover 12
and 10 of the 12 tests, respectively. Outside the range, they cover none of the 8.
The conformal guarantee holds on average over the calibration distribution and says
nothing about a shifted input, whereas the certificate holds for each output.

Learned-majorant methods for neural solutions \cite{fanaskov2024,fanaskov2026}
evaluate a functional majorant \cite{repin2008}. For an admissible potential $v$ and
any current $y$ with no flux through the insulating walls, where $y$ need not be
divergence-free, it states that
$\lVert\sigma\nabla(v-u)\rVert\le\lVert\sigma\nabla v-y\rVert+C_F\,\sigma_{\min}^{-1/2}\lVert\nabla\cdot y\rVert$,
the latter an unweighted $L^2$ norm.
Every vertical line through the monopolar domain ends on the electrode or the ground
at both ends, so the Friedrichs constant
$C_F=H/\pi=\SI{31.8}{mm}$, with $H=\SI{100}{mm}$ the height of the domain, holds
rigorously. We recover $y$ from
the potential by nodal averaging of $\sigma\nabla v$ and set its normal component to
zero on the axis and the shaft, which is the least expensive choice, and we
evaluate the majorant in double precision. With this current, the majorant gives no
finite upper bound on the impedance for any surrogate here, nor for our solver's own
solution. On that solution, its flux term, $1.45\times10^{-5}$, is close to
$\eta=1.30\times10^{-5}$, but its Friedrichs term is $3.1\times10^3$ against a field
norm of 0.10. Of $\lVert\nabla\cdot y\rVert^2$, 99\% lies in the two \SI{1}{nm}
cells at the electrode--shaft junction, where the potential behaves like the square
root of the distance, and even the cells more than \SI{1}{mm} from both corners give
a Friedrichs term of $1.5\times10^{-3}$, a hundred times $\eta$. A learned,
optimized, or locally equilibrated current \cite{braess2009} would lower this term,
but only an exactly divergence-free current removes it, and with such a current, the
majorant reduces to the hypercircle bound used here.

\section{Discussion}\label{sec:discussion}

The experiments separate two questions that a mesh study conflates, namely
whether the discretization converges and whether it converges to the solution of the
stated problem. A mesh study answers the first question, within its asymptotic
assumptions and with judgment about safety factors. Because it compares a solver only
with itself, it cannot answer the second. Five plausible bugs and an LLM-written
solver that never solves the equation all converged, as a correct solver does. The
certificate answers both questions for the run at hand, and its width reports the
quality of that run. For conventional solvers, the two methods are therefore
complementary, with a mesh study for exploration and a certificate for the runs that
support a decision. For learned surrogates, the certificate applies to any output and
is informative when the fields are accurate in energy on the certificate grid. Networks
that predict POD coefficients, and a neural operator whose output lies on a mesh
without long thin cells, gave such fields. The same operator with free nodal outputs
on the certificate grid did not, despite small nodal errors. On this problem, moving
the operator's output onto the triangle mesh narrowed its bounds from $10^3$ times
the impedance to a few percent, and twice as many Fourier modes then brought them below
1\%, whereas more training members did not.

A user supplies the specification, a tensor grid whose lines contain the breakpoints
of the geometry and of the tiles, and the nodal values of the potential, and
of the stream function if the solver computes one. A solver on its own
mesh evaluates its potential at the nodes of the grid and gives the boundary nodes
their exact values, and the adapter then costs about as much as a primal solve. The
vectorized check of \cite{dovelos2026} takes 4--15\% of our solver's time for both
fields and about 30\% of FEniCSx's \SI{3.8}{s}. A surrogate built for the native route outputs both
fields on the grid, so its certificate costs one forward pass and one check. An
operator with output on another mesh adds a transfer, which takes \SI{0.02}{s} here.

The route also determines which bounds come from the solver. Solvers that return
both fields, as FEniCSx did here and as the reduced-order surrogates do by
construction, supply both bounds themselves and give the strongest and cheapest
evidence.

A certificate bounds numerical error only; whether the stated model describes
tissue is a question of validation \cite{dovelos2026}. The test problem is linear,
steady, elliptic, and axisymmetric. The hypercircle relies on the first three
properties, whereas axisymmetry only reduces the problem to two dimensions.
Certificates for a solver on its own mesh include the error of transferring its field
to the certificate grid, which a checker for the solver's own elements would remove.
Tile bounds use the global $\eta$, so local quantities are certified less tightly
than a converged mesh study estimates them, and localized bounds are the main open
improvement. Three LLM-written programs and five plausible bugs are a case study, too
few for statistics. The neural operators were not tuned beyond the number of Fourier
modes, with one training run per variant. The variants with more training members and
more Fourier modes were run after the first trained operator had been evaluated on
the same 20 tests, and all variants were selected on the calibration members only.
On this two-dimensional problem, a certified prediction of the operator is only
about \SI{0.4}{s} faster than a certified FEniCSx solve on the same mesh, whose
interval is 2--3$\times10^5$ times narrower. A substantial time saving from certified surrogates
would therefore have to come from larger problems, in which a solve is expensive
while the check remains linear in the size of the grid. The checker, our solver, and the adapter share one implementation of the
specification's boundary-node rules, so an error there would be common to all
three. The vectorized checker is rigorous under the assumptions of \cite{dovelos2026}.

In three dimensions, the current
can be supplied as the curl of a N\'ed\'elec vector potential, which keeps it
exactly divergence-free in floating point. A conductivity that increases with the field keeps the energy convex,
and convex duality then gives certificates of the same kind \cite{repin2008}.
Surrogates can be made to emit admissible fields with distance-function
constructions \cite{sukumar2022}, and a vectorized checker over tetrahedra would make
meshes of millions of elements practical to check.

\section{Conclusion}
The checker of \cite{dovelos2026}, without any change, verified the outputs of
solvers other than our own. It certified FEniCSx from the package's own potential and
current as tightly as our own solver. It caught five
plausible bugs and an LLM-written solver that a mesh study accepted, and it certified
a correct LLM-written solver to 0.1\%. It gave valid bounds for each prediction of
neural networks on a reduced basis, narrow inside their training range and valid
outside it, where conformal intervals failed. It also showed that a neural operator with free
nodal outputs on a grid with cells as thin as \SI{1}{nm} received uninformative bounds
despite small nodal errors. With its output mapped onto a triangle mesh, training in
the energy norm, and twice as many Fourier modes, the same architecture received bounds
narrower than 1\% in all 12 tests inside its training range. With the vectorized checker of \cite{dovelos2026},
a native certificate costs 4--15\% of our solver's time to check, and a
potential-only output also needs the adapter, about one primal solve.

The check needs no expert and no reference solution, and it holds for each output,
wherever its input lies. It is informative only when the output is accurate in energy
on a grid that resolves the problem, and outputs with both fields need no adapter. The next steps are a second test problem, localized bounds, a formally
verified checker, and three-dimensional certificates.

\section*{Code Availability}
The checker, the solvers, the stored LLM-written programs and plausible bugs, and
the scripts that reproduce every number in this paper are available at
\url{https://github.com/kostisdov/certsim} (tag \texttt{v2.1-preprint}) under the
Apache License 2.0.

\end{document}